\documentclass[a4paper,usenatbib]{mem}
\usepackage{natbib}
\usepackage{graphicx}
\usepackage[a4paper]{hyperref}
\begin{document}
\def\teff{$T\rm_{eff }$}
\def\kms{$\mathrm {km s}^{-1}$}
\def\ltsima{$\; \buildrel < \over \sim \;$}
\def\lsim{\lower.5ex\hbox{\ltsima}}
\def\gtsima{$\; \buildrel > \over \sim \;$}
\def\gsim{\lower.5ex\hbox{\gtsima}}
\def\mdot {\dot M}
\newcommand{\be}{\begin{equation}}
\newcommand{\en}{\end{equation}}
\newcommand{\ergs}{\rm \ erg \; s^{-1}}
\def\kms  {\rm \ km \, s^{-1}}
\def\cms  {\rm \ cm \, s^{-1}}
\def\gs   {\rm \ g  \, s^{-1}}
\def\cmtre {\rm \ cm^{-3}}
\def\cmdue {\rm \ cm^{-2}}
\def\gcmdue {\rm \ g \, cm^{-2}}
\def\gcm  {\rm \ g \, cm^{-3}}
\def\rsole {~R_{\odot}}
\def\msole {~M_{\odot}}
\def \mdotav {\langle \dot {M}\rangle }
\def\deg {$^\circ$}
\def\ppm{$\pm$}

\title{Neutron star observations with WFXT}

   \subtitle{}

\author{
S. \,Campana\inst{1} 
          }

  \offprints{S. Campana}

\institute{
Istituto Nazionale di Astrofisica --
Osservatorio astronomico di Brera, Via E. Bianchi, I-23807,
Merate (LC), Italy
\email{sergio.campana@brera.inaf.it}
}

\authorrunning{S. Campana}

\titlerunning{NS observations with WFXT}

\abstract{The Wide-Field X-ray Telescope (WFXT) is a proposed NASA mission dedicated to performing 
surveys of the sky in the soft X-ray band ($0.3-6$ keV). The key characteristics of this missions are 
a constant point spread function with Half Energy Width of $\sim 5$ arcsec over $\sim 1$ degree field of view
as well as an effective area $\sim 10$ times larger than the one of Chandra.
Despite the fact that the mission is tailored for extragalactic purposes, we show here that extremely 
interesting results can also be obtained on the study of neutron stars.
}
\maketitle{}

\section{Introduction: status}

Neutron stars are formed in supernova explosions and live their early life
as rotationally powered emitters, shining mainly in the high energy band. A small fraction 
of the neutron star spin-down power goes in the radio band in the form of pulsed emission, 
making their discovery possible. 
As newborn objects, neutron stars are also very hot (millions of degrees) and emit 
in the soft X--ray band thanks to the cooling of the compact object.
As the neutron star ages its spin-down power and internal heat decreases 
and it becomes readily unobservable.
Only for compact objects in close binary systems there is an additional way to power their
emission thanks to the exchange of mass from the companion to the neutron star.
Accretion of matter onto a compact object naturally leads to emission in the X--ray band,
powering the so-called X--ray binaries.

Stable mass transfer onto compact objects produces the brightest objects in the X--ray sky.
For this reason our knowledge of the population of X--ray binaries 
in the Galaxy started with the first all-sky hard-band survey from the Uhuru satellite.
Monitoring instruments such as the RossiXTE ASM and now INTEGRAL, Swift BAT and MAXI provide a nearly
real-time census of the population of bright X--ray binaries in our Galaxy and in our closeby neighboorhood.
With these instruments we have access, however, only to the brightest tip of the population.
It was clear from the first X--ray missions that, together with persistent sources,
there is a large population of transient X--ray binaries which spend most of their time 
($90-99\%$) in quiescence and show signs of X--ray activity only for very limited periods of time
(during which they share the same properties of persistently bright sources).
With the coming of new facilities such as XMM-Newton and Chandra it became clear that 
intermediate luminosity X--ray binaries are also present, but difficult to discover and, in turn,
difficult to study. 

\begin{figure*}[!tb]
  \resizebox{0.5\hsize}{!}{\includegraphics{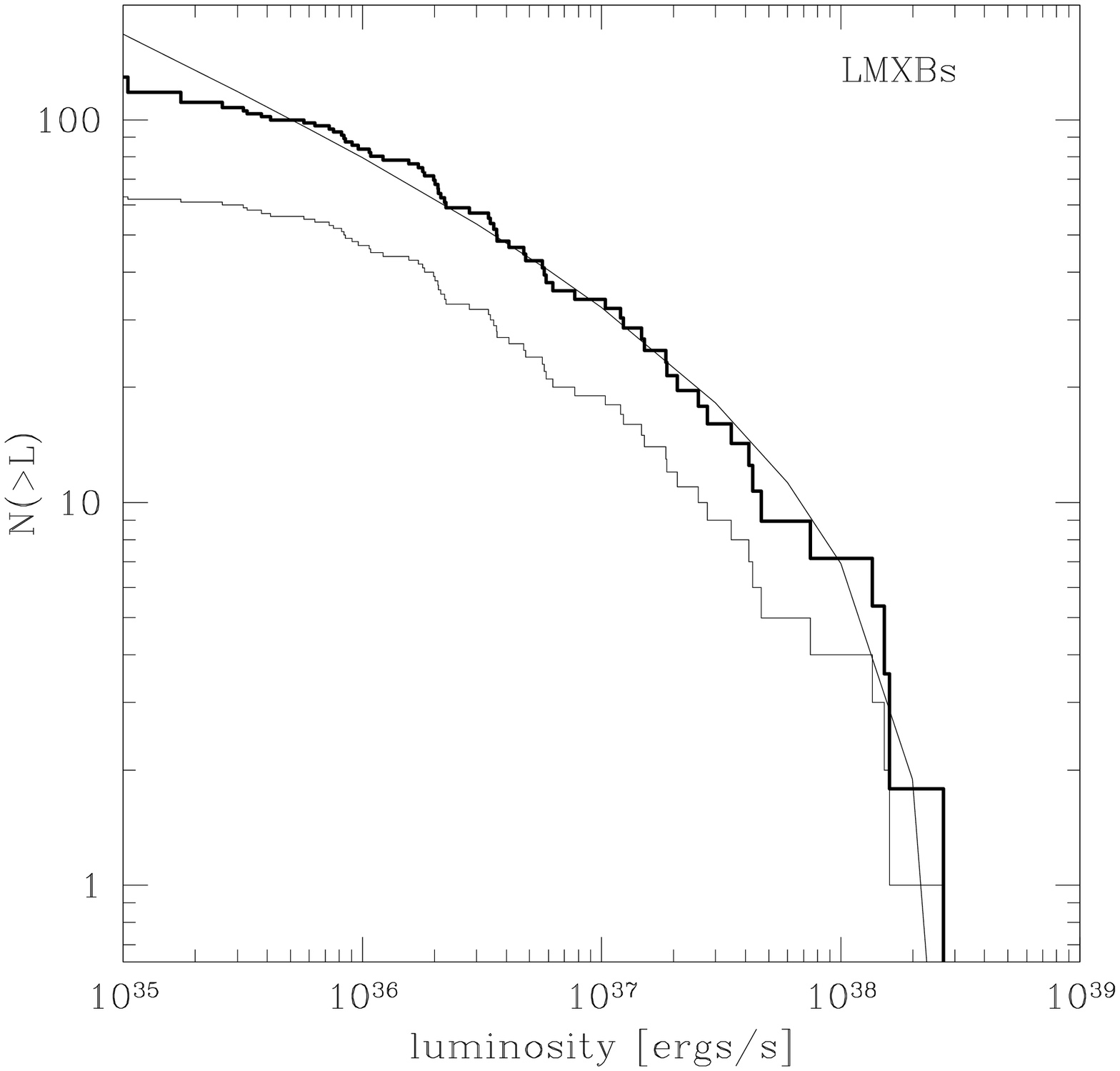}}
  \resizebox{0.5\hsize}{!}{\includegraphics{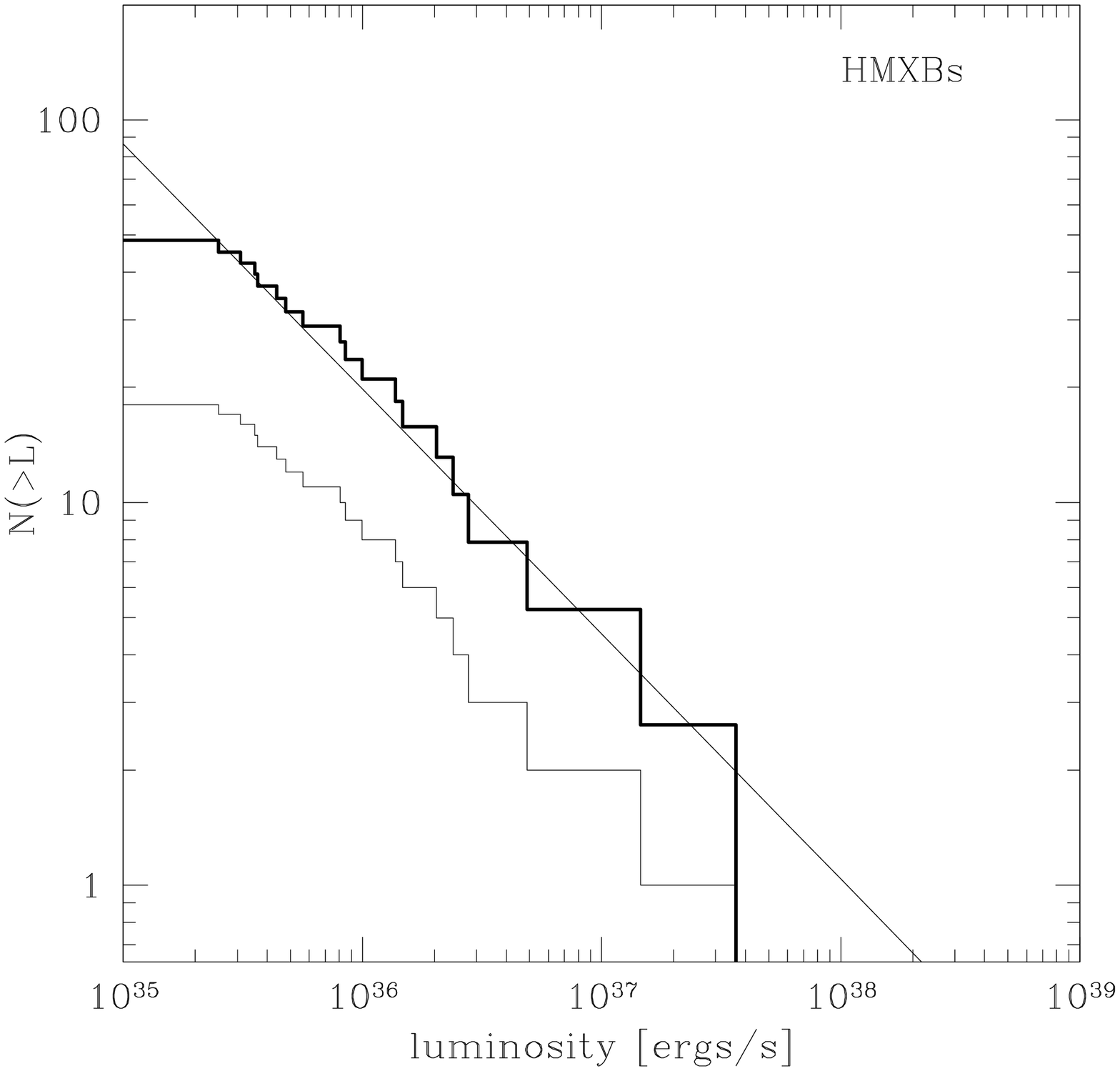}}
  \caption{The apparent (thin histogram) and volume corrected (thick
  histogram) cumulative luminosity function for Low Mass X--ray Binaries and High Mass X--ray Binaries. The
  solid lines are the best fits to the data (from \citep{grimm02}).} 
  \label{fig:lf}
\end{figure*}

Our present view on quiescent and intermediate luminosity X--ray binaries comes from the ROSAT All Sky Survey 
(RASS) in the soft band only and from partial or limited serendipitous surveys carried out with imaging satellites 
like XMM-Newton and Chandra.

\section{A WFXT survey of the Galactic plane: neutron stars}

Actually our knowledge of X--ray binaries as a population relies only on studies 
with the RossiXTE ASM, providing luminosity function of high-mass and low-mass X--ray binaries
(depending on the mass of the companion) down to luminosities of the order of 
$10^{35}-10^{36}$ erg s$^{-1}$ (\citep{grimm02}, see also Fig. \ref{fig:lf}). 
This clearly provides only a biased view of the population missing the great 
majority of faint objects. 
In addition, below this limiting luminosity level accretion onto neutron stars in high mass 
(magnetic field $B\sim 10^{12}$ G and spin periods in the few seconds range for the fastest 
pulsators) and low mass (magnetic field $B\sim 10^{8-9}$ G and spin periods of a few milliseconds) 
X--ray binary transients might enter in accretion regimes different from the 
direct fall of matter onto the neutron star surface (e.g. \citep{campana98}).
These regimes (e.g. propeller, reactivation of a radio pulsar) are basically unexplored as
a population. In quiescence X--ray binary transients are observationally in the $\sim 5\times 
10^{31}-10^{33}$ erg s$^{-1}$ range (e.g. \citep{campana04}).

\begin{figure*}[!tb]
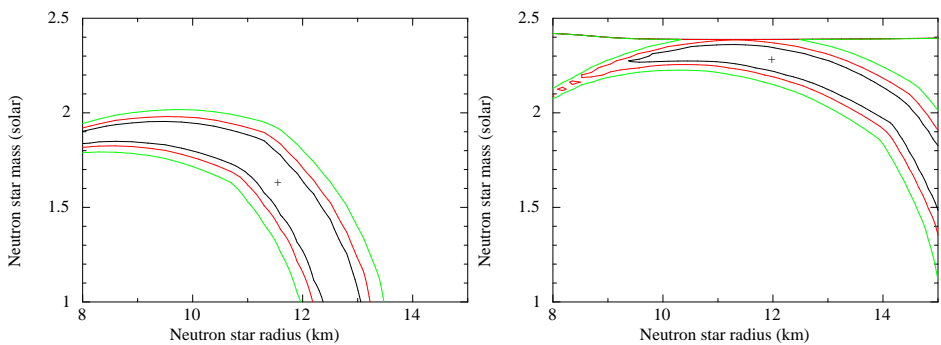

  \resizebox{0.45\hsize}{!}{\includegraphics[angle=270]{ns_mr1.eps}}
  \resizebox{0.45\hsize}{!}{\includegraphics[angle=270]{ns_mr2.eps}}
  \caption{Typical mass-radius relation that can be obtained with WFXT 
observing for 100 ks a neutron star low mass transient in quiescence.
Simulations were carried out taking the quiescent transient in Omega Cen
as a template. In the left panel a mass and radius of $1.66\msole$ and 11.6 km 
were selected, respectively; in the right panel $2.25\msole$ and 12 km.
} 
  \label{fig:ns}
\end{figure*}

A WFXT survey of the Galactic plane comparable in depth with the `wide' survey (i.e. reaching a flux 
limit of $\sim 3\times 10^{-15}$ erg cm$^{-2}$ s$^{-1}$ will reach a luminosity limit of 
$\sim 10^{32}$ erg s$^{-1}$ throughout the Galaxy, providing a complete census of the X--ray binary 
population. A complete census of the X--ray binary population will help constraining the formation and 
evolutionary models. 

A similar mapping can be achieved on the Magellanic Clouds with a survey comparable in depth to the 
`medium' survey. Studying the properties of X--ray binary sources in the Magellanic Clouds 
rather than in our own Galaxy presents several advantages: $i$) the distance of all sources are well known; 
$ii$) the much lower column density allows us to investigate a much wider spectral range than it is
possible in the Galactic plane. The importance of this low column density is highlighted by e.g.
the large number of supersoft sources discovered in the Magellanic Clouds; $iii$) since the metallicities of
the Magellanic Clouds differ from that of our Galaxy, a comparison of their X--ray population will help us
understanding the role of abundances in their properties.

\subsection{Globular Clusters} 

Globular clusters contain a large number of X--ray binaries, that are formed 
thanks to close encounters \citep{heinke03} and the large majority of them 
are quiescent. The X--ray spectrum of a transient low mass X-ray binary in quiescence
comprises two spectral components: one hard usually modelled with a power law (with variable importance 
of a source-by-source basis, from $\lsim 3\%$ to $\sim 50\%$) and 
the other soft modelled with a black body emission. The soft component is also consistent 
with emission coming from the cooling of the entire neutron star surface that has been 
heated during (transient) accretion episodes \citep{brown98}. This emission is well 
understood and, if data of very good quality are gathered, in principle, it can provide a
tool to disentangle the small spectral differences induced by different neutron star masses
and radii. Given the large area of WFXT, 100 ks observation will allow to set strong constraints
on the neutron star equation of state through observations of transient low mass X-ray binaries 
in quiescence (see \ref{fig:ns}).

\begin{figure*}[!tb]
\resizebox{0.95\hsize}{!}{\includegraphics{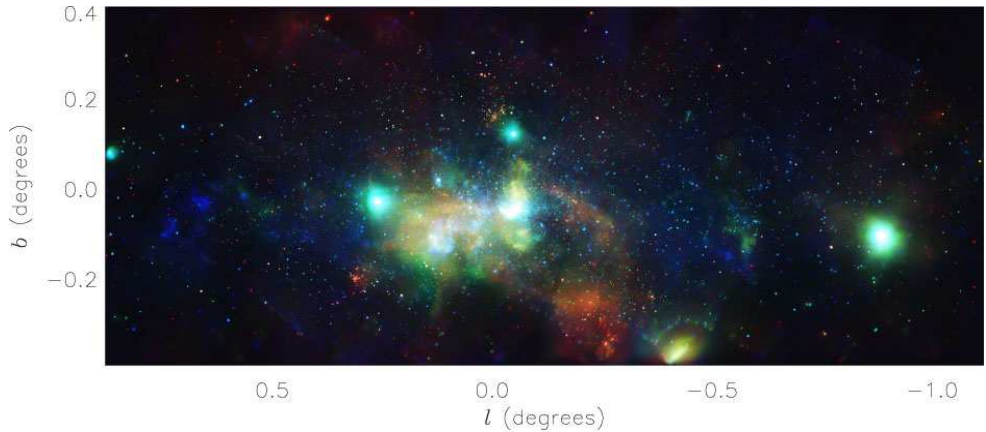}}
\caption{Three-color image of the Galactic center region. Red is 1--3 keV, green is 3--5 keV, and blue is 5--8 keV,
from \citep{muno09}. } 
  \label{fig:gc}
\end{figure*}

In addition, globular clusters contain also a large number of recycled millisecond pulsars 
\citep{bogdanov06}. A statistical study of millisecond radio pulsars can provide insight on the energy 
conversion mechanism of spin down power into high energy photons.

\subsection{Galactic Center} 

Thanks to the Chandra observatory the Galactic center region has been mapped in exquisite details
\citep{wang02,muno09}. A scan of two degree across the Galactic center has been carried out (with 2 Ms exposure) 
reaching a completeness 0.5--8 keV flux limit of $4\times 10^{32}$ erg s$^{-1}$ and up to an order of magnitude more 
sensitive in the deepest exposure around Sgr A$_∗$ (see Fig. \ref{fig:gc}). 
9017 X--ray sources were detected. The majority of the absorbed 
sources ($N_H>4\times 10^{22}$ cm$^{-2}$) are made by cataclysmic variables, even if a number of transients have 
been discovered. WFXT can cover the same area more deeply by an order of magnitude in in 200 ks. 
This opens the possibility of variability studies either temporal and spectral.  
Monitoring programs can be very effective in discovery faint or very faint transients ($L\sim 10^{34}-10^{36}$ 
erg s$^{-1}$) that cannot be detected and followed by all-sky monitor instruments. 
Explorative campaigns have been carried out in the Galactic center region with Chandra and XMM-Newton 
\citep{wijnands06}. These systems are poorly studied and only focussing telescope surveys can reveal 
and study their population \citep{campana09}.

\begin{figure*}[!htb]
	\centering
		\includegraphics[width=0.47\textwidth]{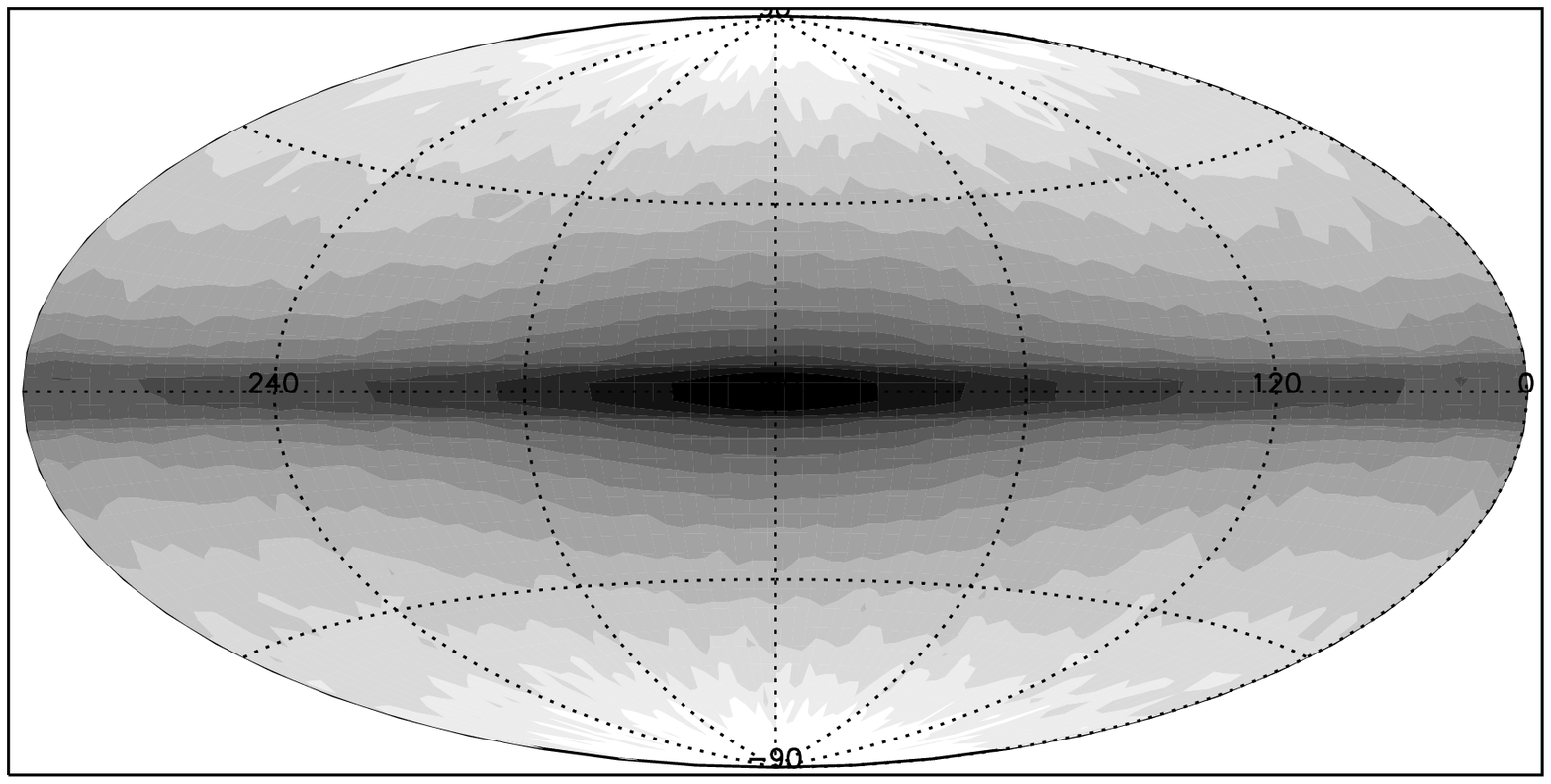}
		\includegraphics[width=0.47\textwidth]{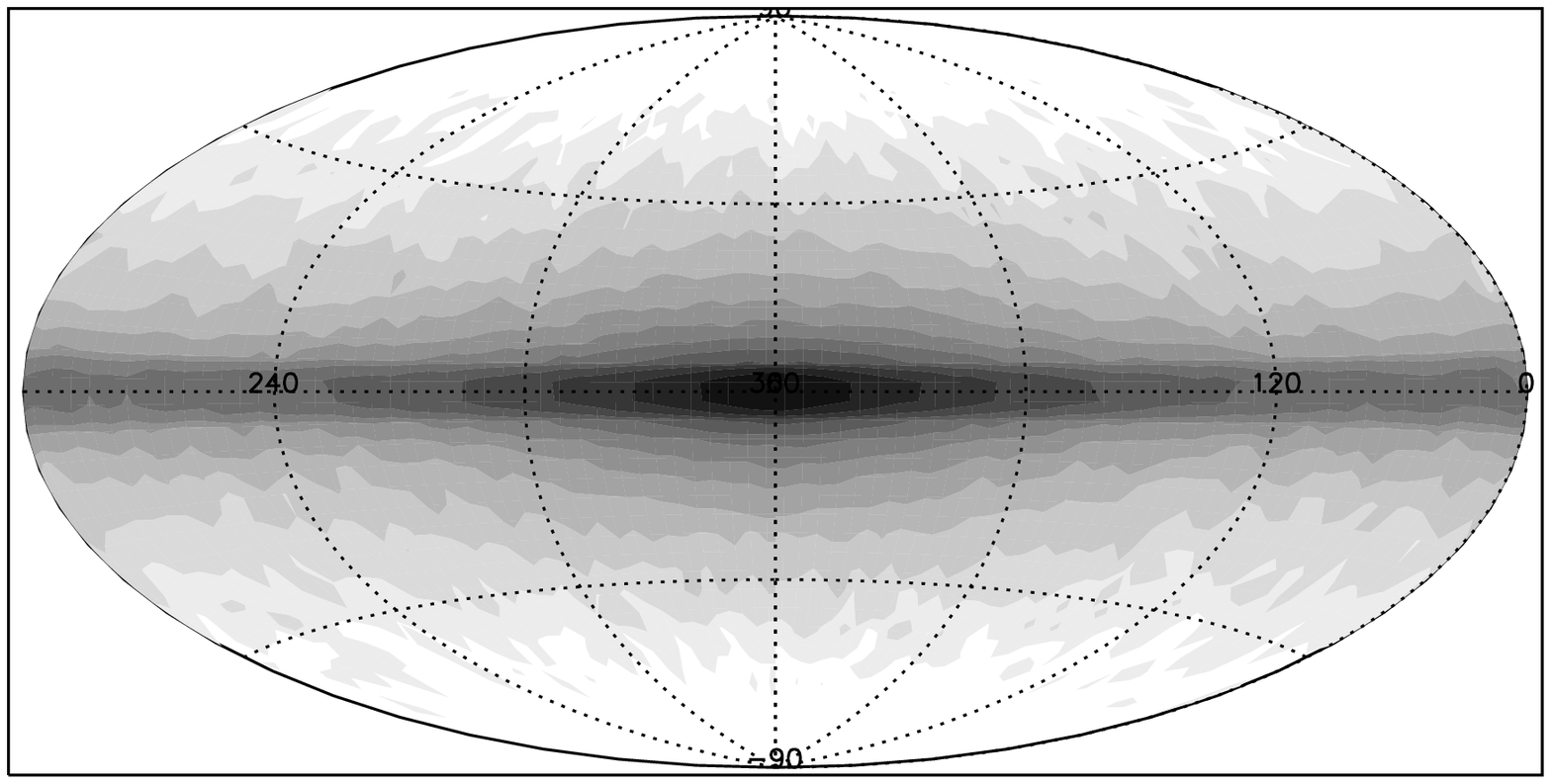}
		\includegraphics[width=0.47\textwidth]{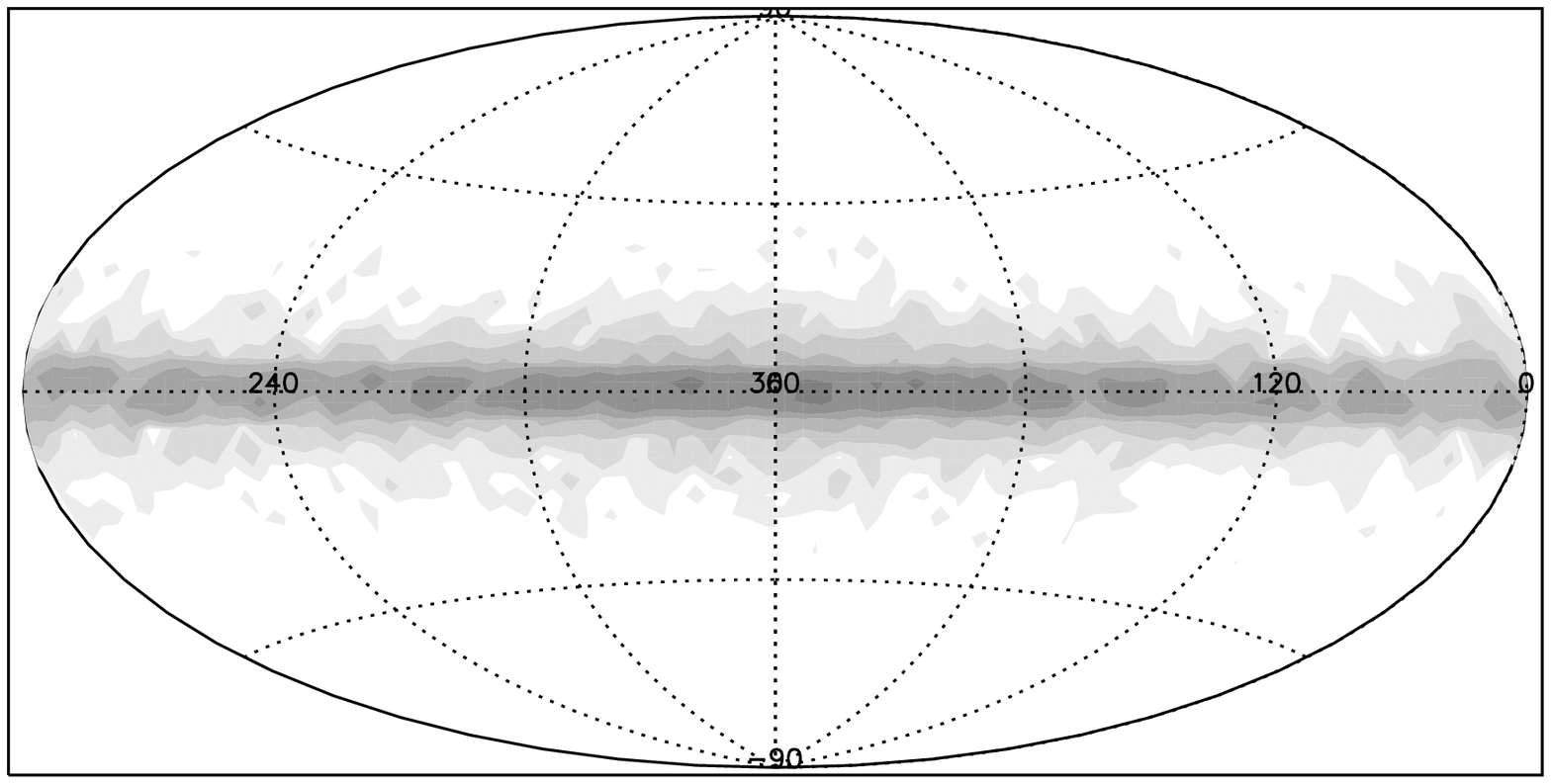}
\caption{Sky maps of the projected density ($N_{star}=10^{9}$) - The cut-off distances are 30 kpc (upper panel), 
10 kpc (central panel) and 3 kpc (lower panel) respectively.
The density scale is normalized to the maximum density at 30 kpc, from \citep{sartone10}.}
	\label{fig:skymaps}
\end{figure*}

\subsection{Old neutron stars} 

About $10^9$ neutron stars are thought to populate our Galaxy, but only $\sim 2\times 10^3$ 
are directly observed as radio pulsars or as accretion-powered X--ray binaries (see Fig. \ref{fig:skymaps}).
In principle also the accretion of the interstellar medium material may make isolated neutron 
stars shine, and their weak luminosity could be detected in soft X--rays. Recent ROSAT observations 
have convincingly shown that neutron stars accreting from the interstellar medium are extremely rare, 
if observed at all, in contrast with earlier theoretical predictions. In addition, accreting objects 
can be confused with much younger, cooling neutron stars. However, a combination of observations
and theoretical modeling may help in discriminating between the two classes \citep{treves00}.

Clearly also isolated cooling neutron stars are extremely important targets since they can shed
light on the supernova explosion rate in the Galaxy and chemical evolution.  
The  ROSAT All-Sky-Survey is the only available survey for this kind of studies.
Turner et al. (2010), using new and archival observations made with the Swift satellite and other 
facilities, examined 147 X--ray sources selected from the RASS Bright Source Catalog (BSC) searching 
for isolated neutron stars (INS). Independent of X--ray spectrum
and variability, the number of INSs is $\lsim ≤48$ ($90\%$ confidence). Restricting attention to 
soft ($T < 200$ eV), non-variable X--ray sources they put an all-sky limit
of $\lsim 31$ INSs. Five new objects were also detected. A future (nearly) all-sky X--ray survey with 
WFXT can be expected to increase the detected population of X--ray-discovered INSs from the 8 to 50 in the BSC, 
to (for a disk population) 240 to 1500, which will enable a more detailed study of neutron star population models.

\section{Conclusions}

The Wide Field X-Ray Telescope (WFXT) is a medium-class mission designed to be 2-orders-of-magnitude 
more sensitive than any previous or planned X--ray mission for large area surveys and to match in 
sensitivity the next generation of wide-area optical, IR, and radio surveys. 
The WFXT mission is scientifically broad. The main focus of the mission is on extragalactic science 
but, as shown above, many important topics can be covered by WFXT concerning neutron stars.

\begin{acknowledgements}
The author would
  like to thank all the members of the WFXT team for a number of
  enlightening discussions.
\end{acknowledgements}

\bibliographystyle{aa}
\bibliography{master}
\end{document}